\documentclass[aps,prl,twocolumn,groupedaddress]{revtex4}

\usepackage{graphicx}
\usepackage{amssymb}

\newcommand{\ybinag}{YbIn${_{1-x}}$Ag${_{x}}$Cu${_{4} }$}
\newcommand{\ybin}{YbInCu${_{4} }$}
\newcommand{\ybag}{YbAgCu${_{4} }$}

\newcommand{\tk}{${T_{K} }$}

\newcommand{\sigone}{$\sigma_{1}(\omega)$}
\newcommand{\EF}{$E_{F}$}

\newcommand{\x}{$x$}

\newcommand{\f}{${f}$}

\newcommand{\s}{\hspace{.05cm}}

\begin{document}


\title{Kondo Scaling in the Optical Response of \ybinag}

\author{Jason N. \surname{Hancock}}

\author{Tim \surname{McKnew}}
\author{Zack \surname{Schlesinger}}
\affiliation{Physics Department, University of California Santa Cruz, Santa Cruz, CA 95064, USA}

\author{John L. \surname{Sarrao}}
\affiliation{Los Alamos National Laboratory, Mail Stop K764, Los Alamos, New Mexico 87545, USA}

\author{Zach \surname{Fisk}}
\affiliation{National High Magnetic Field Laboratory, Tallahassee, Florida 32310, USA}
\affiliation{Department of Physics, Florida State University, Tallahassee, Florida 32306, USA}

\date{\today}
\begin{abstract}
Theoretical work on Kondo systems predicts universality in the scaling of observable quantities with the Kondo temperature, \tk. Here we report infrared-frequency optical response measurements of the correlated system \ybinag. We observe that \x-dependent variations in the frequency and strength of a low energy excitation are related to the \x-dependent Kondo temperature. Comparison of the inferred trends with existing theory and a model calculation provides a framework in which to view these experimental results as scaling phenomena arising from local-moment/conduction electron hybridization.
\end{abstract}
\pacs{}
\maketitle
The study of the Kondo problem has contributed much in the way of
theoretical technology (e.g. renormalization group) as well as insight into the
possible manifestations and phenomenologies of interacting systems with many degrees of
freedom. One unifying effect born from the years of investigation is the
Kondo resonance--a many-body collective
excitation responsible for much of the interesting low energy behavior displayed by
rare-earth and transition metal systems.

\ybag\ provides an example of a moderately heavy fermion system. At low
temperatures, the ${j=\frac{7}{2}}$ Yb moment is screened by conduction
electrons giving an enhanced Pauli paramagnetic susceptibility\cite{cornelius97,sarrao2}.
Enhancement is also found in the
Sommerfeld coefficient ${\gamma\sim 250\hspace{.05cm}mJ/mol K^{2}}$, indicating a
large effective carrier mass\cite{sarrao2}, or equivalently, a large density of states
near the chemical potential. At high temperature, the Yb moment acts independently of the carriers, which show unenhanced mass behavior. The crossover between the high and low temperature regimes occurs around a characteristic scale \tk, which is intimately related to the degree to which the low temperature properties are enhanced. 

Substituting In for Ag gives very different behavior. \ybin\ displays a first-order isostructural electronic phase transition at $T_{\nu}\simeq42\s K$ from a free-moment semimetal to a low-temperature metallic mixed-valent phase\cite{kindler,lawrence2}. This phase transition begins a phase boundary in the ${x-T}$ plane of \ybinag\ which increases in temperature as \x\ is increased and ends in a critical point around ${x\simeq0.3}$. This phase boundary is of considerable interest\cite{lawrence2,kindler,cornelius97,sarrao2,garner00,demsar,freericks} because it involves mainly electronic degrees of freedom and because there is as yet no real understanding of or agreement regarding its fundamental origin. When approaching an understanding, one must address the periodic Anderson model (PAM), which provides a powerful basis for interpreting the phenomenology of heavy fermion and mixed-valent systems, where crossover behavior is the norm.
While much of  the phenomenology of \ybinag\ is consistent with the PAM, the occurrence of an unexpected phase transition at low $x$ could indicate the need for an additional
interaction term in its minimal model Hamiltonian. It is natural in this context to consider whether adding conduction electron interactions (\textit{c.f.} Giamarchi \textit{et al.}\cite{giamarchi93} and Freericks \textit{et al.}\cite{freericks}) which are not generally included in the PAM, could provide competing influences necessary to understand the origin of this phase transition. Further exploration in this area may help establish a connection between the domain of Mott-Hubbard physics, where strong conduction electron interactions lead to a phase transition, and the moment-compensation physics of the Anderson and Kondo models.

In low-temperature \ybin, where the Kondo scale is large, there appears a distinct feature in the mid-infrared conductivity that is not present in the high temperature phase\cite{garner00}. It has been suggested that similar mid-infrared features found in other heavy fermion materials\cite{garner00,dordevic,degiorgi01} are a manifestation of local moment (Kondo lattice/periodic Anderson model) phenomena\cite{coleman, coxcomm,millis87a,millis87b}. Here, by studying the Ag concentration dependence of the dynamical conductivity in \ybinag, we effectively vary \tk\ within the low temperature phase and observe systematic changes in characteristics of the mid-infrared excitation. We then explore the scaling behavior of this excitation in the context of a simple local-moment hybridization picture, and establish both the identification and phenomenology of the optical signature of the Kondo resonance.

\begin{figure}[htbp]
\centerline{\scalebox{.7}{\includegraphics{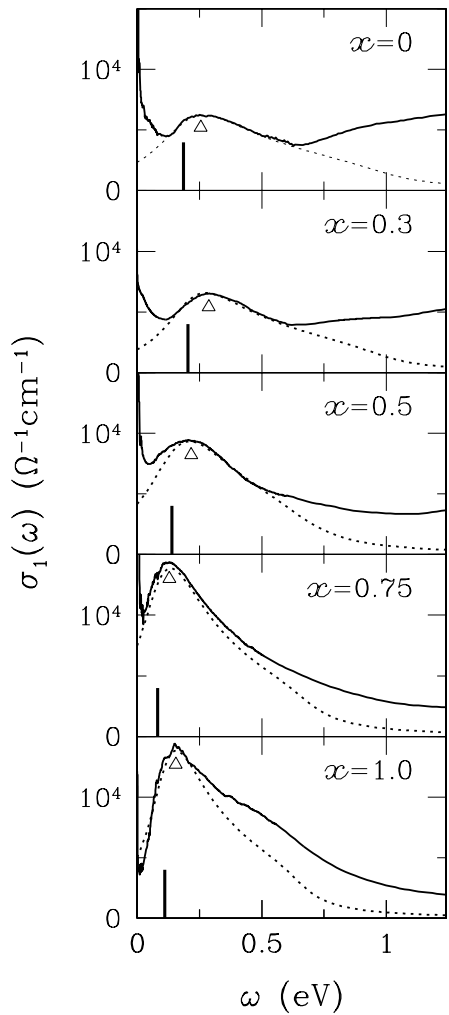}}}
\caption{
The real part of the low-temperature optical conductivity of \ybinag\  (${T = 20 \hspace{.05cm}K}$) is shown for
five values of \x. The open triangles mark peak frequencies. The dotted curves
refer to a PAM-based joint density-of-states calculation in the text and the dark vertical bars
indicate a threshold frequency from that calculation.}
\label{fig:firfig}
\end{figure}

We have measured the reflectivity of single
crystals of a sequence of samples of \ybinag\ with ${x=0}$, 0.3, 0.5, 0.75
and 1.0. Our measurements cover the frequency range from ${5\hspace{.05cm}meV}$ to ${6.2\hspace{.05cm} eV}$
with detailed temperature dependent data taken between ${5 \hspace{.05cm}meV}$ and ${2.8\hspace{.05cm} eV}$. 
A Kramers-Kronig transform is applied to the measured reflectivity to determine the real
and imaginary parts of the dynamical conductivity. For the purposes of the transform,
Hagen-Rubens terminations are used below ${5\hspace{.05cm} meV}$. At high frequency (above ${6\hspace{.05cm} eV}$) each reflectivity spectrum is extrapolated to a common
value of 0.08 at ${\omega = 15\hspace{.05cm} eV}$ 
and then continued as a constant to ${25 \hspace{.05cm}eV}$. Between ${25 \hspace{.05cm}eV}$ and ${50\hspace{.05cm} eV}$
an ${\omega^{-2}}$ form is used for ${R(\omega)}$ and above ${50 \hspace{.05cm}eV}$ the
free-electron form ${\omega^{-4}}$ is assumed.  In order to assure that our essential results below ${1\hspace{.05cm} eV}$ do not depend on the high-frequency extrapolations, 
we have experimented with a number of termination protocols including other common values and coalescence frequencies as well as constant extrapolations above ${6.2\hspace{.05cm} eV}$.  These show convincingly that our results regarding trends in the \x\ and \tk\ dependence of \sigone\ below ${1\hspace{.05cm} eV}$ are not significantly influenced by any of the extrapolations above ${6.2\hspace{.05cm} eV}$.

In Figure \ref{fig:firfig}, the low temperature conductivity is shown for the five \x\ values studied. 
A prominent feature centered around ${\frac{1}{4}eV}$ for \x=0  (\ybin) changes only slightly as \x\ is increased to 0.3, however
further doping to \x=0.5 and 0.75 causes both a
red shift and strengthening of this feature.  The triangles mark the frequency of the peak of \sigone\ and the vertical bars mark the location of a threshold frequency obtained from a model calculation described below.

\begin{figure} \centerline{\scalebox{.5}
{\includegraphics{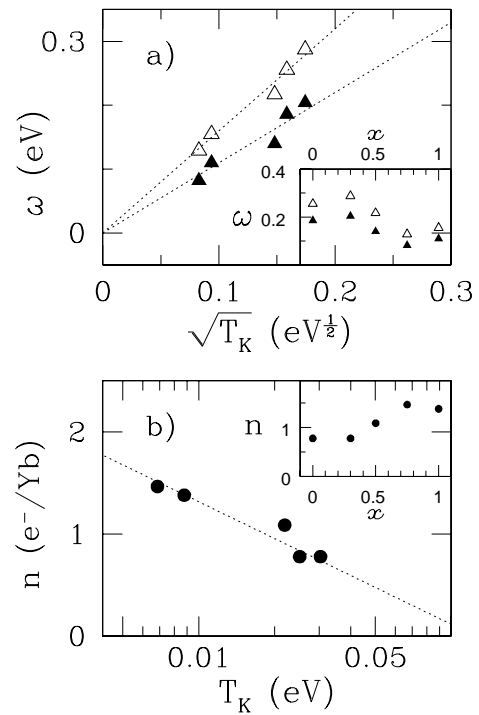}}}
\caption{a) Characteristic frequencies, ${\omega_{\triangle}}$ (${\triangle}$) and ${\omega_{th}}$
(${\blacktriangle}$) are shown as a function of \x\ (inset) and the Kondo temperature, \tk. b) Spectral weight is shown versus \x\ (inset) and \tk. \tk\ is obtained from low-temperature magnetic susceptibility\cite{cornelius97} measurements. Spectral weight is calculated using a bare band mass ${m=4m_{e}}$. Dotted lines show fits to the data based on model equations in the text.}
\label{fig:ntw}
\end{figure}

In Figure 2a these two characteristic frequencies, corresponding respectively to the peak and threshold of this excitation, are plotted both as a function of \x\ and as a function of \tk. 
Values of \tk\ as a function of \x\ are obtained from measurements of magnetic susceptibility\cite{cornelius97}. 
The complex dependence on \x\ becomes monotonic when plotted vs \tk.
The dotted curves in Figure 2a refer to a theoretical scaling described below.

In Figure 2b an integral of  \sigone\ in the vicinity of this excitation is  shown as a function of \x\ and \tk.
This spectral weight, or strength, is defined by the integral,
\begin{equation} n(\omega)=\frac{2 m}{\pi
e^{2}}\int_{0^{+}}^{\omega}\sigma_1(\omega^{\prime})\mathrm{d}\omega^{\prime}.
\label{eqn:sw} \end{equation}
where the limits of integration, ${6\hspace{.05cm} meV}$ and ${0.5\hspace{.05cm} eV}$, encompass the ${\frac{1}{4}\hspace{.05cm}eV}$  excitation and exclude a very small Drude peak at very low frequency.
Neither integration limit is critical; in fact a lower limit of 0 and upper limits between anywhere between  ${0.4\hspace{.05cm} eV}$ and ${1\hspace{.05cm} eV}$ produce the same \x\ dependence which makes one confident that the ${n}$ versus \x\ dependence shown here is an essential characteristic of the data, and independent of any of the details of the choices we have made in the analysis. 

Our discussion of the data centers on the PAM dispersion relations illustrated in the inset of Fig. 3a which address the essential physics of hybridization of conduction electrons with local moments and the appearance of the Kondo resonance at \EF. At energies far from the chemical potential, the upper and lower bands, ${\epsilon^{+}}$ and ${\epsilon^{-}}$, follow closely the unrenormalized free carrier dispersion. Near \EF, the Fermi surface opens (and the bands flatten) to accommodate the \f-electron weight projected up to the Fermi level as a result of hybridization. This reorganization of the bands in the vicinity of the Fermi level
is due to many-body interactions, the strength of which is characterized by the parameter ${\tilde{V}}$. The resultant narrow peak in the density of states, called the Kondo, or Abrikosov-Suhl, resonance is central to the understanding of heavy fermion and mixed-valent phenomenology\cite{degiorgi99,hewson, jarrell95}.

The hybridization-induced splitting creates the
possibility of vertical transitions from filled states below \EF\ to
unoccupied levels above \EF, as illustrated by the arrows of Figure 3a. The threshold for these transitions occurs at a frequency ${\omega=2\tilde{V}}$, where ${\tilde{V}}$ is the
hybridization strength renormalized by the on-site \f-electron
repulsion. This energy scales with the Kondo temperature as\cite{coleman, coxcomm,millis87a,millis87b,grewe84}
\begin{equation}\tilde{V}=\sqrt{T_{K} B}\label{eq:tkb}\end{equation}
where ${B}$ is related to the conduction electron bandwidth\cite{millis87a}.
At threshold, the nesting of the coherent(lower) and incoherent(upper) bands leads to a very high joint density of states for vertical transitions and hence a strong peak in the conductivity,
which we show below can be related with the width of the Kondo resonance, \tk.

\begin{figure} \centerline{ \scalebox{.5}{\includegraphics{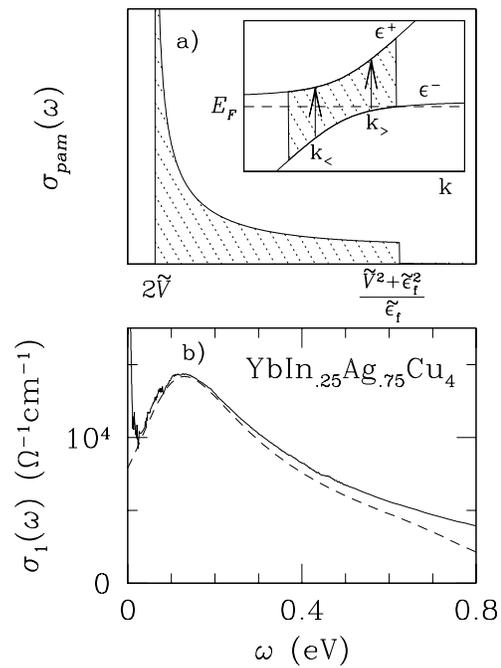}}}
\caption{The inset shows the PAM dispersion relations with arrows
indicating transitions made allowed by the existence
of the Kondo resonance.  a) shows $\sigma_{pam}(\omega)$ calculated from those transitions.
b) shows a model conductivity calculated using ${\tilde{V}=41\hspace{.05cm} meV}$, ${\tilde{\epsilon_{f}}=2.5\hspace{.05cm}meV}$,
${E_{F}=1\hspace{.05cm}eV}$, ${|\textbf{p}_{+,-}|^{2} k_{F}=4.673\hspace{.05cm}\AA^{-3}}$ and ${\Delta=0.12\hspace{.05cm}eV}$.}
\label{fig:jdos}
\end{figure}

To proceed with the analysis, we use the Kubo-Greenwood formula\cite{dressel}:
\begin{equation}
\sigma_{1}(\omega) = \frac{\pi e^{2}}{m^{2}\omega}
\sum_{\ell,\ell^{\prime}}JDOS_{\ell,\ell^{\prime}}(\omega)|\textbf{p}_{\ell,\ell^{\prime}}|^{2}
\label{eq:conduct1}
\end{equation}
where ${|\textbf{p}_{\ell,\ell^{\prime}}|}$ denotes the dipole matrix element
connecting electronic bands ${\ell}$ and ${\ell^{\prime}}$, and
${JDOS_{\ell,\ell^{\prime}}(\omega)}$ is the corresponding joint density of states. Applying this formula to hybridizing quasiparticles as though they were electrons allows an exploration of the phenomena of the mid-infrared conductivity in the context of the PAM. In that case there are two relevant bands, $\epsilon^{+}$ and $\epsilon^{-}$ and our model approach leads to:
\begin{equation}
\sigma_{pam}(\omega) = \frac{e^{2}}{4 \pi^{2} m^{2}\omega}
\int_{\Delta\epsilon=\omega}\frac{dS}{|\nabla_\textbf{k}(\epsilon^{+}-\epsilon^{-})|}|\textbf{p}_{+,-}|^{2}
\label{eq:conduct2}
\end{equation}
where ${|\textbf{p}_{+,-}|}$ is the matrix element for the one remaining term in the $JDOS_{\ell,\ell^{\prime}}$ sum of Eq. (\ref{eq:conduct1}).

For each value of ${\omega>2\tilde{V}}$, there are two
contributions to the integral: one originating from levels
inside the unrenormalized Fermi surface (${k_{<}}$); the other from the states
occupied as a result of renormalization, i.e., outside the unrenormalized
Fermi surface (${k_{>}}$). An example of two such transitions with the same $\omega$ are indicated by vertical arrows in the inset of
Figure \ref{fig:jdos}.

Simplifying equation (\ref{eq:conduct2}) by making an isotropic approximation and the approximation of a constant $|\textbf{p}_{+,-}|$ (discussed further below), one obtains
\begin{equation}
\sigma_{pam}(\omega) = \frac{e^{2}|\textbf{p}_{+,-}|^{2}}{4 \pi^{2} m^{2}\omega}
\sum_{k'=k_< ,k_>}\frac{4 \pi k^{2}}{|\partial_{k}(\epsilon^{+}-\epsilon^{-})|}\Bigg{|}_{k=k'}
\label{eq:conduct3}
\end{equation}
Model spectra can then be calculated using the explicit PAM dispersion relation\cite{grewe84,hewson},
\begin{equation}
\label{eqn:epm}
\epsilon^{\pm}=\frac{E_{F}+\tilde{\epsilon_{f}}+\epsilon_{\textbf{k}}\pm \sqrt{(E_{F}+\tilde{\epsilon_{f}}-\epsilon_{\textbf{k}})^{2}+4 \tilde{V}^{2}}  }{2},
\end{equation}
where $E_F$ is the Fermi level, $\tilde{V}$ is the renormalized hybridization strength, introduced above, and ${\tilde{\epsilon_{f}}}$ is the \f-level position renormalized by
on-site \f-electron repulsion which defines the scale of the low-energy physics and is commonly identified with the impurity
Kondo temperature, \tk.

An example of such a calculation for ${\tilde{V}=41\hspace{.05cm} meV}$ and ${\tilde{\epsilon_{f}}=2.5\hspace{.05cm}meV}$, is shown in Figure 3a. The onset for optical transitions appears as a cusp at ${2\tilde{V}}$\footnote{A second feature of the theoretical $\sigma_{pam}(\omega)$ occurs at a higher frequency ${\omega=(\tilde{V}^{2}+\tilde{\epsilon_{f}}^{2})/\tilde{\epsilon_{f}}}$, and results from the initial state energy exceeding \EF, where no occupied levels are available. This high energy scale is of order the bandwidth and is used as the upper limit of integration for determining the strength.}. While the strength of this cusp is an important feature associated with nesting, the extreme sharpness is an artifact of assuming infinite lifetimes for the quasiparticle states. We rectify this by convoluting this model result with a Lorentzian of constant width to \textit{post hoc} emulate the finite-lifetime effects. The magnitude of the broadening parameter, $\Delta$, gives an estimate of the quasiparticle lifetime. The values used here of around $0.12\s eV$, correspond to a lifetime of order $\tau=\hbar/\Delta\sim6\s ps$, consistent with pump-probe lifetime measurements of \ybag\cite{demsar}.

In Fig. 3b we present this calculated $\sigma_{pam}(\omega)$ together with our measured \sigone\ for \x=0.75. Based on the quality of this fit, the PAM parameters ${\tilde{V}=41\hspace{.05cm} meV}$ and ${\tilde{\epsilon_{f}}=2.5\hspace{.05cm}meV}$ can be associated with this composition. With ${\tilde{V}}$ and $\tilde{\epsilon_{f}}$ as adjustable parameters, we have similarly modeled the mid-infrared conductivity \sigone\ for each \x\ value as shown by the dashed curves in Figure \ref{fig:firfig}. The vertical bars mark the threshold frequencies $\omega_{th}=2\tilde{V}$.
These threshold frequencies are plotted versus \x\ and \tk\ in Figure 2a.  The dotted curve through those data shows the consistency between our experimental result and the theoretically based scaling of Eq. \ref{eq:tkb}.

With regard to the \textit{total strength}, one can obtain an analytic result in our approach by setting the two sphere areas in (\ref{eq:conduct3}) equal to ${4\pi k_{F}^{2}}$. This approximation avoids the effects of bare band structure details while including the influence of the strong cusp at ${2\tilde{V}}$. Solving for the wavevectors ${k_{>}}$ and ${k_{<}}$ using the condition ${\epsilon^{+}-\epsilon^{-}=\omega}$, one can obtain a model strength,
which is essentially the shaded area of Fig. 3a, of:
\begin{eqnarray}
n_{pam} & = & \frac{2m}{\pi e^2}\int_{2\tilde{V}}^{\frac{\tilde{V}^{2}+\tilde{\epsilon_{f}}^{2}}{\tilde{\epsilon_{f}}}}\sigma_{pam}(\omega) d\omega   \\
& \simeq & \frac{4|\textbf{p}_{+,-}|^{2} k_{F}}{\pi^2}\ln\Big(\frac{\tilde{V}}{\tilde{\epsilon_{f}}}\Big)\\
& \simeq & \frac{4|\textbf{p}_{+,-}|^{2} k_{F}}{\pi^2}\ln\Big(c\sqrt{\frac{B}{T_{K}}}\Big)\label{eqn:answer},
\end{eqnarray}
where $c$ is defined by $T_K=c\s \tilde{\epsilon_f}$. This logarithmic scaling behavior is compared to the data in Figure 2b.

It is important to realize that the simple arguments leading up to equation (\ref{eqn:answer}) are not strictly rigorous and in fact careful analysis reveals that a $\textbf{k}$-independent hybridization leads to a $|\textbf{p}_{+,-}|$ which depends on $\textbf{k}$, and therefore $\omega$, in a manner that will reduce the conductivity at high frequency\cite{coxcomm}. One expects that in a real system, transitions between Cu-In-Ag $d$ orbital bands and Yb \f-band states are likely to be allowed due to their relative spatial displacement.  Incorporation of this possibility into theory can be done by allowing a dispersive $\tilde{V}_\textbf{k}$, which would tend to restore some of the high frequency conductivity that is underestimated in our simplified approach. A more sophisticated approach to $n(T_K)$ would be valuable in the ongoing effort toward a quantitative understanding of strongly correlated electrodynamics.

Equations (9) and (2) do, however, qualitatively explain the observed dependence of the frequency and strength on \tk.   In addition these equations show a interesting relationship between the scaling of  $n(T_K)$ and  $\omega_{th}(T_K)$.
Using the impurity model result for $c$\footnote{For ${N=2j+1=8}$, the Fermi liquid theory of the Anderson impurity model predicts\cite{hewson} ${\tilde{\epsilon_{f}}=k_{B}T_{K} N^{2} 2 \sin(2\pi/N)/w_{N}\pi(N-1)\simeq 1.517\hspace{.05cm} T_{K}}$.}, and fitting the measured $n(T_K)$ to Eq. (\ref{eqn:answer}), one gets a value for the band parameter $B=0.36\s eV$. This value is determined solely by the extrapolation of the data to the \tk\ axis, and so does not depend on $k_F$ or the matrix elements present in the coefficient of the logarithm. One can independently obtain a value for $B$ from fitting the \tk\ dependence of the excitation frequency (Figure 2a), $\omega_{th}=2\tilde{V}=2\sqrt{T_K B}$, which leads to $B=0.30 \s eV$. The observation that these independently obtained values of $B$ are comparable shows a non-trivial relationship between the scaling behavior of strength and characteristic frequency in our optical data. 
Overall our results indicate that \tk\ scaling is present in the low temperature finite-frequency dynamics of \ybinag\ and can be addressed in the context of local-moment models. Further work may be directed toward greater understanding of this low-$T$ scaling behavior as well as unexplained temperature dependent behavior of \ybinag.

\begin{acknowledgments}
We would like to thank D. L. Cox for helpful discussions regarding the theory of Kondo systems and have also benefitted from conversations with D. N. Basov, A. L. Cornelius, P. A. Lee, B. S. Shastry, and A. P. Young. We also gratefully acknowledge S. L. Hoobler and Y. W. Rodriguez for technical assistance. Work at UCSC supported by NSF Grant Number DMR-0071949. ZF acknowledges support of NSF Grant Number DMR-0203214.
\end{acknowledgments}

\bibliography{zs-short,ybcu,valence,kondo03}

\end{document}